\documentclass[twocolumn,showpacs,amssymb,floatfix]{revtex4}
\usepackage{graphicx}
\usepackage{dcolumn}
\usepackage{bm}
\begin{document}
\title{Gaussian random waves in elastic media}
\author{Dmitrii N. Maksimov$^1$, Almas F. Sadreev$^{1,2}$}
\address{1) Institute of Physics, Academy of Sciences, 660036 Krasnoyarsk,
Russia\\ 2) Department of Physics and Measurement Technology,
Link\"oping University, SE-581 83 Link\"oping, Sweden}
\date{\today}

\begin{abstract}
Similar to the Berry conjecture of quantum chaos we consider
elastic analogue which incorporates longitudinal and transverse
elastic displacements with corresponding wave vectors. Based on
that we derive the correlation functions for amplitudes and
intensities of elastic displacements. Comparison to numerics in a
quarter Bunimovich stadium demonstrates excellent agreement.
\end{abstract}

 \maketitle

\section{Introduction}

Attracting interest in the field of wave chaos \cite{stockmann},
elastomechanical systems are being studied analytically,
numerically, and experimentally. Weaver first measured the few
hundred lower eigen frequencies of an aluminum block and worked
out the spectral statistics \cite{weaver1}. Spectral statistics
coinciding with random matrix theory were observed in experiments
for monocrystalline quartz blocks shaped as three-dimensional
Sinai billiards \cite{ellegard2}, as well as, in experimental and
numerical studies of flexural modes \cite{legrand,schaadt}  and
in-plane modes \cite{ellegard4,ellegard5} for stadium-shaped
plates. Statistical properties of eigen functions describing
standing waves in elastic billiards were first reported by Schaadt
{\it et al.} \cite{ellegard6}. The authors measured the
displacement field of several  eigen modes of a thin plate shaped
as a Sinai stadium. Due to a good preservation of up-down symmetry
in the case of thin plates they dealt with two types of modes. The
flexural modes with displacement perpendicular to the plane of the
plate are well described by the scalar biharmonic Kirchoff-Love
equation \cite{graaf,landau}. In this case a good agreement with
theoretical prediction for both intensity statistical distribution
and intensity correlation function was found. However in the case
of in-plane displacements described by the vectorial Navier-Cauchy
equation \cite{graaf,landau} an agreement between the intensity
correlator experimental data and the theory was not achieved
\cite{ellegard6}.

The aim of present letter is to present an analogue of the Berry
conjecture for elastic vibrating solids and derive the amplitude
and intensity correlators with corresponding comparison to
numerics. Quite recently, Acolzin and Weaver suggested a method to
calculate the intensity correlator of vibrating elastic solids
\cite{akolzin}. Based on the Green's function averaging technique
they succeeded to derive the intensity correlator of flexural
modes generalized due the finite thickness of a plate. Although
the method might be used for the in-plane modes in elastic chaotic
billiards, we propose here a more simple and physically
transparent approach based on random superposition of traveling
plane waves (Gaussian random wave (GRW) or the Berry function
\cite{berry}). We show that the approach allows to derive all
kinds of the correlators of RGW not only in infinite elastic media
but also to take into account the double ray splitting at the
boundary of a plate that plays a significant role in the
elastomechanical chaotic motion \cite{tanner,couchman}. We
restrict ourselves to the two-dimensional case because of the
current experiments available. Note however that the method can be
easily generalized for the three-dimensional case.

\section{Analogue of the Berry conjecture in elastic media}
Shapiro and Goelman \cite{shapiro1} first presented statistics of
the eigenfunctions in chaotic quantum billiard although their
numerical histogram was not compared with the Gaussian
distribution. This was done by McDonnell and Kaufmann
\cite{McDonald} who concluded that the majority ($\approx 90\%$)
of the eigenfunctions of the Bunimovich billiard are a Gaussian
random field. Later it was confirmed by numerous numerical and
experimental studies. The simple way to construct RGF is random
superposition of particular solutions of Eq. (\ref{Helm})
\cite{rayleigh,berry,ebeling} with sufficient number $N$. Thus we
come to the Berry conjecture in the form \cite{pnini,stockmann}
\begin{equation}\label{berryQM}
\psi_B({\bf x})=\sqrt{\frac{1}{N}}\sum_{n=1}^N\exp[(i(\theta_n+\bf
{k}_n{\bf x})],
\end{equation}
where the phases $\theta_n$ are random distributed uniformly in
range $[0, 2\pi)$ and all the amplitudes are taken to be equal
(one could assume random independent amplitudes, without any
change in the results). The wave vectors ${\bf k}_n$ are uniformly
distributed on a d-dimensional sphere of radius $k$. It follows
now from the central limit theorem that both $Re\psi_B$ and
$Im\psi_B$ are independent Gaussian variables. In a closed
billiard the Berry function is viewed as a sum of many standing
waves, that is simply real or imaginary part of function
(\ref{berryQM}).

In our case one has to construct a RGW-function describing
acoustic in-plane modes. These modes are described by a
two-dimensional Navier-Cauchy equation \cite{landau,achenbach}
\begin{equation}\label{basic}
  \mu \nabla^2 {\bf u}+(\lambda+\mu)\nabla(\nabla {\bf u})+\rho\Omega^2 {\bf
  u}=0
\end{equation}
where ${\bf u}(x,y)$ is the displacement field in the plate,
$\lambda, \mu$ are the material dependent Lam$\rm\acute{e}$
coefficients, and $\rho$ is the density. Introducing elastic
potentials $\psi$ and ${\bf A}$ with the help of the Helmholtz
\cite{achenbach} decomposition the displacement field ${\bf u}$
could be written,
\begin{equation}\label{poten}
  {\bf u}={\bf u}_l+{\bf u}_t, ~~{\bf u}_l=\nabla\psi, ~{\bf
  u}_t=\nabla\times{\bf A}
\end{equation}
Eq. (\ref{basic}) reduces to two Helmholtz equations for the
elastic potentials
\begin{eqnarray}\label{Helm}
  -\nabla^2\psi=k_l^2\psi, \nonumber\\
  -\nabla^2{\bf A}=k_t^2{\bf A}.
\end{eqnarray}
Here $k_l=\omega/c_l, ~k_t=\omega/c_t$ are the wave numbers for
the longitudinal and transverse waves, respectively and
$\omega^2=\rho \Omega^2/E$, where $E$ is Young's modulus. In the
two-dimensional case potential {\bf A} has only one none-zero
component $A_z$ and the dimensionless longitudinal and transverse
sound velocities $c_{l,t}$ are given by
\begin{equation}\label{velocities}
c_l^2=\frac{1}{1-\sigma^2}, ~c_t^2=\frac{1}{2(1+\sigma)},
\end{equation}
where $\sigma$ is Poisson's ratio \cite{landau, achenbach}. $E$
and $\sigma$ are functions of the Lam$\rm\acute{e}$ coefficients
\cite{landau, achenbach}. Our conjecture is that both elastic
potential be statically independent Berry-like functions
(\ref{berryQM}). We write the potentials in the following form
\begin{eqnarray}\label{berry}
\psi({\bf x
})=\frac{a_l}{ik_l}\sqrt{\frac{1}{N}}\sum_{n=1}^N\exp[i({\bf
k}_{ln} {\bf x}+\theta_{ln})],\nonumber \\ A_z({\bf x
})=\frac{a_t}{ik_t}\sqrt{\frac{1}{N}}\sum_{n=1}^N\exp[i({\bf
k}_{tn}{\bf x}+\theta_{tn})],
\end{eqnarray}
where $\theta_{ln}$, $\theta_{tn}$ are  statistically independent
random phases. The wave vectors ${\bf k}_{ln}$ and ${\bf k}_{tn}$
are uniformly distributed on circles of radii $k_l$ and $k_t$
respectively.  According to (\ref{poten}) the components $u, v$ of
the vectorial displacement field ${\bf u}$  could be now written
\begin{eqnarray}\label{full}
u({\bf x})= \sqrt{\frac{1-\gamma}{N}}\sum_{n=1}^N
  \cos\phi_{ln}\exp[i({\bf k}_{ln}{\bf x}+\theta_{ln})]\nonumber\\
  +\sqrt{\frac{\gamma}{N}}\sum_{n=1}^N
  \sin\phi_{tn}\exp[i({\bf k}_{tn}{\bf x}+\theta_{tn})],\nonumber\\ v({\bf x})=
\sqrt{\frac{1-\gamma}{N}}\sum_{n=1}^N
  \sin\phi_{ln}\exp[i({\bf k}_{ln}{\bf x}+\theta_{ln})]\nonumber\\
  -\sqrt{\frac{\gamma}{N}}\sum_{n=1}^N
  \cos\phi_{tn}\exp[i({\bf k}_{tn}{\bf x}+\theta_{tn})],\nonumber\\
\end{eqnarray}
where $\phi_{ln}, ~\phi_{tn}$ are the angles between ${\bf
k}_{ln}, ~{\bf k}_{tn}$ and the x-axis respectively. The
prefactors $a_l=\sqrt{\gamma}, a_t=\sqrt{1-\gamma}$ are chosen
from the normalization condition $ \langle {\bf u}^\dag {\bf u}
\rangle = 1 $, and $\langle \ldots\rangle$ means average over the
random phase ensembles. Parameter $\gamma$ ranges from 0 (pure
transverse waves) to 1 (pure longitudinal waves). By similar way
one can construct the elastomechanical  GRW for a closed system.
One can see that the Berry analogue of chaotic displacements
(\ref{full}) is not a sum of two independent GRWs (or two
independent Berry functions) $a_l\psi_l+a_t\psi_t$ as it was
conjectured by Schaadt {\it et al.} \cite{ellegard6} with
arbitrary coefficients $a_l, ~a_t$. In fact, each component $u$
and $v$ in Eq (\ref{full}) is related to the Berry functions
(\ref{berry}) via space derivatives in accordance to relations
(\ref{poten}).

\section{Correlation functions}
First, we calculate the amplitude correlation functions in chaotic
elastic plate for in-plane GRW (\ref{full}). For quantum
mechanical GRW (\ref{berryQM}) the two-dimensional correlation
function $$ \langle \psi_B({\bf x+s})\psi_B^{*}({\bf
x})\rangle=J_0(s)$$ was found firstly by Berry \cite{berry}.
Straightforward procedure of averaging over ensembles of random
phases $\theta_{ln}, ~\theta_{tn}$ and next, over angles of
k-vectors gives
\begin{figure}[t]
\includegraphics[scale=0.4]{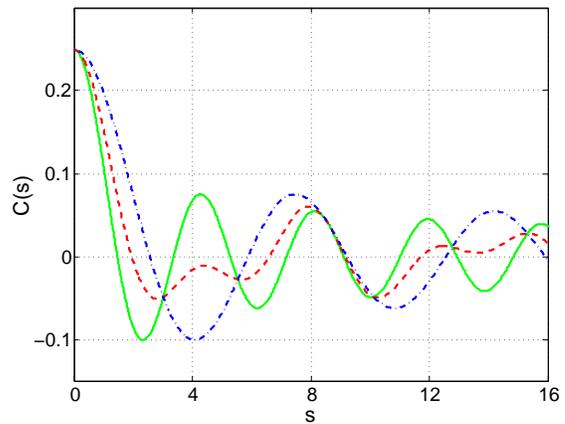}
\caption{Correlation function (\ref{correliso}) for $\gamma=0$
(green solid line), $\gamma=0.5$ (red dashed line), and $\gamma=1$
(blue dot-dashed line); $\sigma=0.345$ (aluminum).}
\label{correlam}
\end{figure}
\begin{eqnarray}\label{correl}
&\langle u({\bf x + s})u({\bf
x})\rangle=\frac{\gamma}{2}({\cos}^2\alpha~f(k_l
s)+{\sin}^2\alpha~g(k_l s))+&\nonumber\\
&\frac{1-\gamma}{2}({\sin}^2\alpha ~f(k_t s)+{\cos}^2\alpha ~g(k_t
s)),&\nonumber\\ &\langle v({\bf x + s})v({\bf
x})\rangle=\frac{\gamma}{2}({\sin}^2\alpha ~f(k_l
s)+{\cos}^2\alpha ~g(k_l s))+&\nonumber\\
&\frac{1-\gamma}{2}({\cos}^2\alpha ~f(k_t s)+{\sin}^2\alpha ~g(k_t
s)), &\nonumber\\ &\langle u({\bf x + s})v({\bf x})\rangle=
{\sin2\alpha}(\frac{1-\gamma}{2}J_2(k_ts)-\frac{\gamma}{2}
J_2(k_ls)),
\end{eqnarray}
where
\begin{equation}
  f(s)=J_0(s)-J_2(s), g(s)=J_0(s)+J_2(s).
\end{equation}
It is important to note that the correlation functions
(\ref{correl}) were obtained for given direction of the vector
${\bf s}$ where $\alpha$ is the angle included between vector $\bf
s$ and the x-axis. However, averaged over all directions of ${\bf
s}$, the first two correlators  simplify
\begin{eqnarray}\label{correliso}
&C(s)=\overline{\langle u({\bf x+s})u({\bf
x})\rangle}=\overline{\langle v({\bf x+s})v({\bf
x})\rangle}&\nonumber\\ &=\frac{1-\gamma}{2}J_0(k_ls)+
\frac{\gamma}{2}J_0(k_ts)&,
\end{eqnarray}
while the third vanishes $\overline{\langle u({\bf x+s})v({\bf
x})\rangle}=0$. One can see that in the averaged case the
amplitude correlation function is defined by two scales because of
two different sound velocities $c_l, ~c_t$; that is obvious. The
correlation function $C(s)$ is shown in Fig. \ref{correlam}.

Next, we calculate the intensity correlation functions
$P(s)=\langle I{\bf x+s})I({\bf x})\rangle$ where the intensity
$I=|{\bf u}|^2$ proportional to the elastic energy of the in-plane
oscillations. In quantum mechanics this value is analogous to the
probability density, the correlation function of which was
calculated by Prigodin {\it et al.} \cite{prigodin}. For the
in-plane chaotic GRW of the form $a_l\psi_l+a_t\psi_t$ Schaadt
{\it et al.} \cite{ellegard6} derived the intensity correlation
function as
\begin{equation}\label{P2shaadt}
  P(s)=1+2[a_l^2J_0(k_ls)+a_t^2J_0(k_ts)]^2.
\end{equation}

Our calculations similar to those as for the amplitude correlation
functions (\ref{correl}) give the different result
\begin{eqnarray}\label{P2}
 & P(s)=1+\frac{1}{2\eta}[(\gamma J_0(k_ls)+(1-\gamma)J_0(k_ts)]^2\nonumber\\
& +\frac{1}{2\eta}[(\gamma J_2(k_ls)-(1-\gamma)J_2(k_ts)]^2&
\end{eqnarray}
where $\eta=1$ for real GRW and $\eta=2$ for complex one. Although
the first term in (\ref{P2}) corresponds to (\ref{P2shaadt}) there
is a different term consisted of the Bessel functions $J_2$. The
mathematical origin of deviation is that formula (\ref{full})
contains the contributions of the components of the wave vectors
${\bf k}_l$ and ${\bf k}_t$ via space derivatives.
\section{Wave conversion at boundary}
Waves propagate freely inside the billiard, that is, the
longitudinal and transverse components are decoupled. Wave
conversion occurs at the boundary according to Snell's law
\begin{equation}\label{Snell}
c_{l}\sin(\theta_{t})=c_{t}\sin(\theta_{l}),
\end{equation}
The reflection amplitudes for each event of the reflection can be
easily found following the procedure described in \cite{landau}.
At first we consider more easy case of the Dirichlet BC (the
boundary is fixed). Approximating the boundary as the straight
lines for the wavelengths much less than the radius of curvature
we have for the reflection amplitudes
\begin{eqnarray}\label{t11}
  t_{ll}=\frac{\cos(\theta_{t})\cos(\theta_{l})-\sin(\theta_{t})\sin(\theta_{l})}
  {\cos(\theta_{t})\cos(\theta_{l})+\sin(\theta_{t})\sin(\theta_{l})},\nonumber\\
t_{lt}=\frac{2\sin(\theta_{l})\cos(\theta_{l})}
  {\cos(\theta_{t})\cos(\theta_{l})+\sin(\theta_{t})\sin(\theta_{l})},\nonumber\\
  t_{tl}=\frac{2\sin(\theta_{t})\cos(\theta_{t})}
  {\cos(\theta_{t})\cos(\theta_{l})+\sin(\theta_{t})\sin(\theta_{l})},\nonumber\\
t_{tt}=\frac{\cos(\theta_{t})\cos(\theta_{l})-\sin(\theta_{t})\sin(\theta_{l})}
  {\cos(\theta_{t})\cos(\theta_{l})+\sin(\theta_{t})\sin(\theta_{l})}.
\end{eqnarray}
\begin{figure}[t]
\includegraphics[scale=0.5]{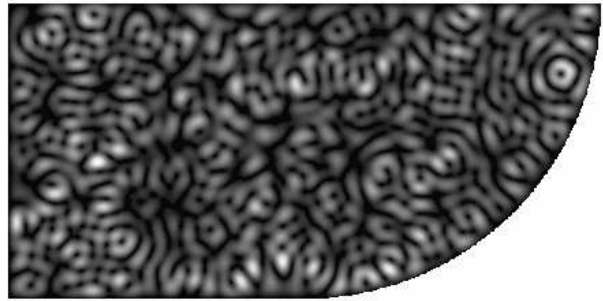}
\caption{The intensity $I=|{\bf u}|^2$ of  the eigen state at
frequency $\omega= 28.4$ in the quarter of the Bunimovich billiard
with fixed boundary, $\sigma=0.345$.} \label{eig}
\end{figure}
Next, we assume that all directions of waves are statistically
equivalent. Then we have for the energy density of reflected wave
\begin{equation}\label{refl}
  \rho_{out}=\gamma(\overline{T}_{ll}+\overline{T}_{lt}) +
  (1-\gamma)(\overline{T}_{tt}+\overline{T}_{tl}),
\end{equation}
where
$$\overline{T}_{ij}=\frac{1}{\pi}\int^{\pi}_{0}t^{2}_{ij}d\theta_i,
~i=l,t.$$ Substituting into here (\ref{t11}) one can obtain after
elementary calculations
\begin{eqnarray}\label{Tout}
&\overline{T}_{ll}=1-\frac{c_t}{c_l}I_1,
~\overline{T}_{lt}=I_2,&\nonumber\\
&\overline{T}_{tt}=1-\frac{2}{\pi}\arcsin\frac{c_t}{c_l}+
\left(\frac{c_t}{c_l}\right)^3I_1,&\nonumber\\
&\overline{T}_{tl}=\frac{2}{\pi}\arcsin\frac{c_t}{c_l}-\left(\frac{c_t}{c_l}\right)^2I_2.&
\end{eqnarray}
We do not present here integrals $I_1, I_2$, since after
substitution of (\ref{Tout}) into (\ref{refl}) they cancel each
other. The equality $\rho_{in}=1=\rho_{out}$ gives a very simple
evaluation
\begin{equation}\label{gamma}
\gamma=\frac{c^{2}_{t}}{c^{2}_{t}+c^{2}_{l}}.
\end{equation}
The next remarkable result is that although the reflection
amplitudes for the free BC \cite{landau} have the form different
from (\ref{t11}) (see, for example, formulas in \cite{landau}) ,
the evaluation of $\gamma$ by the same procedure gives the same
form as for the fixed BC. Therefore, we can conclude that the
result does not depend on either the free BC or the fixed BC is
applied. Using (\ref{velocities}) formula (\ref{gamma}) could be
written in a more simple form
\begin{equation}\label{gam-sigma}
  \gamma=\frac{1-\sigma}{3-\sigma}.
\end{equation}
\begin{figure}[t]
\includegraphics[scale=0.5]{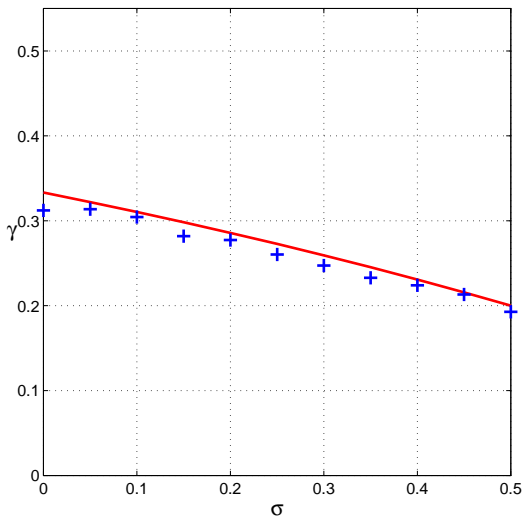}
\caption{Cross marks show numerical results for ratio $\gamma$
averaged over 200 eigenfunctions of the quarter Bunimovich plate.
The solid line is plotted by formula (\ref{gam-sigma}).}
\label{sigma}
\end{figure}
\section{Numerical results and conclusion}
For numerical tests we took the quarter of the Bunimovich billiard
and calculated the eigenstates of the Navier-Cauchy equation
(\ref{basic}) with the fixed BC: $u=0, ~v=0$ at the boundary of
the billiard by the finite-difference method. An example of the
eigenstate in the form of intensity $ I=|{\bf u}|^2$ is presented
in Fig. \ref{eig}. First of all we verified formula
(\ref{gam-sigma}). For each value of Poisson's ratio $\sigma$ in
the range $[0, 0.5]$ with the step 0.05, 200 eigenfunctions of the
billiard were found to calculate averaged $\gamma$. The resulted
dependence of $\gamma$ on $\sigma$ is shown in Fig. \ref{sigma},
that demonstrates a  good agreement with formula
(\ref{velocities}).
\begin{figure}[h]
\includegraphics[scale=0.45]{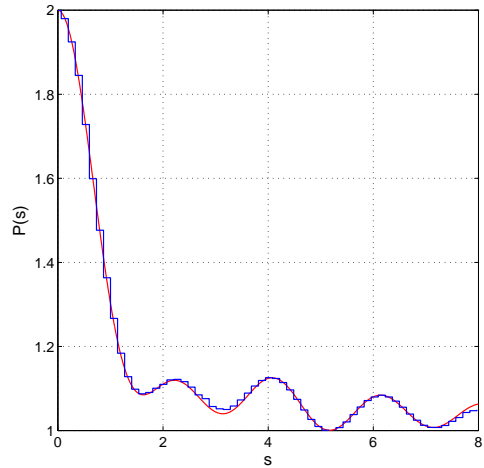}
\caption{The intensity correlation function (\ref{P2}) compared to
the numerics for the same situation as in Fig. \ref{eig}.}
\label{P2comp}
\end{figure}
Thereby we can evaluate $\gamma$ for specific $\sigma=0.345$ which
correspond to aluminum plate and plot the correlation functions.
The intensity correlation function (\ref{P2}) is shown in Fig.
\ref{P2comp} compared to the numerics calculated for the eigen
function presented in Fig. \ref{eig}. One can see a good
coincidence of the theory with the numerical results that
demonstrates correctness of the GRW-approach (\ref{full}) to chaos
in elastic billiards.

{\bf Acknowledgments}. The authors acknowledge discussions with
K.-F. Berggren.  This work is supported by RFBR grant 07-02-00694.


\end{document}